\documentclass[aps,prl,10pt,twocolumn,superscriptaddress]{revtex4-2}
\usepackage{amsmath}
\usepackage{amssymb}
\usepackage[unicode=true,colorlinks=true,citecolor=blue,urlcolor=blue]{hyperref}
\usepackage{graphicx}

\begin{document}

\title{Topology of the generalized Brillouin zone of one-dimensional models}
\author{Heming Wang}
\affiliation{Department of Electrical Engineering, Stanford University, Stanford, California 94305, USA}
\affiliation{Edward L. Ginzton Laboratory, Stanford University, Stanford, California 94305, USA}
\author{Janet Zhong}
\affiliation{Edward L. Ginzton Laboratory, Stanford University, Stanford, California 94305, USA}
\affiliation{Department of Applied Physics, Stanford University, Stanford, California 94305, USA}
\author{Shanhui Fan}
\affiliation{Department of Electrical Engineering, Stanford University, Stanford, California 94305, USA}
\affiliation{Edward L. Ginzton Laboratory, Stanford University, Stanford, California 94305, USA}
\affiliation{Department of Applied Physics, Stanford University, Stanford, California 94305, USA}

\begin{abstract}
The generalized Brillouin zones (GBZs) are integral in the analysis of non-Hermitian band structures.
Conventional wisdom suggests that the GBZ should be connected, where each point can be indexed by the real part of the wavevector, similar to the Brillouin zone.
Here we demonstrate rich topological features of the GBZs in generic non-Hermitian one-dimensional models.
We prove and discuss a set of sufficient conditions for the model to ensure the connectivity of its GBZ.
In addition, we show that the GBZ can become disconnected and have more connected components than the number of bands, which results from the point-gap features of the band structure.
This novel GBZ topology is applied to further demonstrate a counterintuitive effect, where the line gap of an open-boundary spectrum with sublattice symmetry may be closed without changing its point-gap topology.
Our results challenge the current understanding of bands and gaps in non-Hermitian systems and highlight the need to further investigate the topological effects associated with the GBZ including topological invariants and open-boundary braiding.
\end{abstract}
\date{\today}
\maketitle

Band structures are important tools for understanding the behavior of periodic systems \cite{bernevig2013topological, bansil2016colloquium, hasan2019colloquium, cayssol2021topological}, and their scope has been greatly expanded by the introduction of non-Hermitian effects \cite{ashida2020non-hermitian, wang2023non-hermitianREV, wang2021topologicalREV, bergholtz2021exceptional, ding2022non-hermitian, okuma2023non-hermitian, zhang2022review, wang2024non-hermitian, lin2023topologicalREV, gong2018topological, gohsrich2025non-hermitian, zhang2025non-hermitian}. For non-Hermitian band structures, the generalized Brillouin zone (GBZ) replaces the conventional Brillouin zone (BZ) and provides a basis for the modes available in the system \cite{yao2018edge, yokomizo2019non-bloch, yang2020non-hermitian}. For single-band models, it has frequently been assumed that the GBZ behaves like the BZ in terms of its topology, which is a single loop that encircles $z=0$ in the complex $z$ plane \cite{zhang2020correspondence, yang2020non-hermitian, chen2023real}. Furthermore, it has been conjectured \cite{zhang2020correspondence} that the GBZ remains connected for single-band models.

Here, we investigate the connectivity of the GBZ in more detail. In particular, we show that a single-band GBZ can have two or more connected components. This feature is associated with disconnected regions with a winding number of $0$ as found in the band windings of the model and, as such, can be considered a topological effect. We first prove a set of sufficient conditions that ensure the connectivity of the GBZ. These criteria prevent single-band models with a coupling range less than three from possessing a disconnected GBZ. Then, we provide some examples of single-band models possessing disconnected GBZs. Finally, the multiple connectedness of the OBC spectrum is used to demonstrate a process of closing the line gap of the open-boundary spectrum of two-band sublattice-symmetric models without closing its point gap. The closed line gap makes it difficult to integrate the Berry curvature over a single band \cite{zhou2025topological}, and we discuss alternative ways to characterize the topological invariant of such models.

{\it The generalized Brillouin zone.} --
Throughout this Letter, a ``model'' refers to a one-dimensional, tight-binding lattice system with translational symmetry. We will focus on single-band models, where general results will be presented and proved, and discuss specific examples of two-band models towards the end. The band structure of a single-band model in the wavevector space without boundary conditions can be written as
\begin{equation}
E = \sum_{-p\leq j \leq q} t_jz^j
\label{eq:model}
\end{equation}
where $E$ is the state energy, $t_j$ is the strength of the coupling from lattice site $n+j$ to site $n$, $z \equiv \exp(ik)$ with $k$ the complex wavevector and $p$ ($q$) represent the coupling ranges to the right (left) \cite{wang2024non-hermitian}. To avoid degeneracy problems, we always assume that $p\geq 1$, $q\geq 1$, and $t_{-p}t_q \neq 0$.

In the context of non-Hermitian band structures, the coupling coefficients in Eq. (\ref{eq:model}) can be arbitrary complex numbers. As a result, the spectrum of the system is heavily dependent on the boundary conditions imposed on the system \cite{guo2021exact}. For periodic boundary conditions (PBC), the wavevector is real ($\text{Im}\ k=0$, $|z|=1$) and belongs to the conventional BZ. The corresponding spectrum is a closed curve that may enclose regions of nonzero area on the complex energy plane, a feature known as band winding. For open boundary conditions (OBC) with a sufficiently long chain, the wavevector for the spectrum can be
complex and the spectrum can be different from the PBC spectrum, demonstrating the
non-Hermitian skin effect. It can be proved that the OBC spectrum of a single-band model must be connected \cite{ullman1967problem}. Usually, the OBC spectrum is shaped like a tree, where arcs are joined at their endpoints, and may possess rich features in terms of their graph topology \cite{tai2023zoology, xiong2024graph}.

\begin{figure}
\centering
\includegraphics[width=85mm]{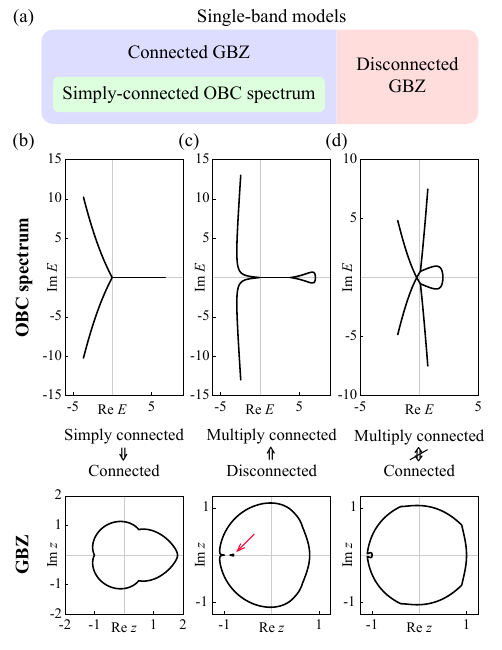}
\caption{Connectivity of the GBZ.
(a) Euler diagram showing that, for single-band models, a simply-connected OBC spectrum implies a connected GBZ, per Theorem I.
(b) The model $E = 5z^{-2} -z^{-1} +5z -z^2$ has a simply connected OBC spectrum (top) and a connected GBZ (bottom).
(c) The model $E = -2z^{-3} +5z^{-2} -z^{-1} +5z -z^2 +2z^3$ has a multiply connected OBC spectrum (top) and a disconnected GBZ (bottom).
(d) The model $E = -2z^{-3} +3z^{-2} -z^2 +2z^3$ has a multiply connected OBC spectrum (top) but a connected GBZ (bottom).
}
\label{fig:1}
\end{figure}

For single-band models, the condition that a complex energy $E$ is on the OBC spectrum is $|z_p|=|z_{p+1}|$, where $z_j(E)$ are the solutions of $z$ to Eq. (\ref{eq:model}) sorted by amplitude in non-decreasing order. The collection of such $z_p$ and $z_{p+1}$ values forms the GBZ. In addition to describing the OBC spectra, the GBZ is also an important concept in understanding non-Hermitian band structures. For example, it is used as integration contours for the non-Hermitian topological invariants \cite{yokomizo2019non-bloch, yao2018edge, zhongwang2025topological, verma2024topological, yang2020non-hermitian, imura2019generalized, hou2022deterministic, verma2024non, liu2023topological} and the Green's functions and dynamics of the model \cite{hu2023greens, liu2023stable, li2022exact, zirnstein2021exponentially, chen2024formal, huang2025complex, xue2025non, xue2021simple, fu2023anatomy, mao2021boundary}. The assumption of a monotonically increasing $\arg(z)$ for GBZ contours is also used in the study of braiding of OBC bands \cite{wojcik2020homotopy, hu2021knots, li2021homotopical, fu2024braiding, wang2025observing, li2022topological, shi2024studies}. In all such cases, it is necessary to consider the topological properties of the GBZ itself. However, no detailed study of the GBZ topology and its implications has been carried out except the conjecture that all single-band GBZs are connected \cite{zhang2020correspondence}, which we will disprove with counterexamples.

{\it Connectivity of the GBZ.} --
We now present the central result of this Letter, which links the connectedness of the GBZ to the simple connectedness of the OBC spectrum [Fig. \ref{fig:1}(a) and \ref{fig:1}(b)]:

{\bf Theorem I.} For a single-band model, if the OBC spectrum of the model is simply connected, then the GBZ associated with the model is connected.

Here, an OBC spectrum is simply connected if it ``has no holes'': every path in the spectrum can be continuously shrunk to a point without leaving the spectrum. An OBC spectrum that is not simply connected is termed multiply connected. The proof proceeds by introducing the inside (outside) of the GBZ, which represents states that are more localized to the left (right) end compared to the OBC states. Under the given conditions, these two regions on the Riemann sphere of $z$ must be simply connected, and the GBZ is the only boundary between them. A more rigorous statement and a complete proof can be found in Supplementary Materials.

Theorem I implies that a model with a disconnected GBZ must possess a multiply connected OBC spectrum [Fig. \ref{fig:1}(c)]. The converse of Theorem I does not hold, and the GBZ of a multiply-connected OBC spectrum may remain connected [Fig. \ref{fig:1}(d)].

The condition in Theorem I requires the OBC spectrum, which may be difficult to compute with high precision to obtain its connectivity properties. We now relate the connectivity of the GBZ to the point-gap properties of the PBC spectrum. Within the point-gap topology, a PBC spectrum is topologically trivial with respect to a reference point $E_0$ if the winding number $w$ of the PBC spectrum with respect to $E_0$ is $0$, and is topologically nontrivial otherwise \cite{kawabata2019symmetry, gong2018topological, okuma2020topological, zhang2020correspondence}. The PBC point-gap topology is useful in determining the existence of non-Hermitian skin effects and its robustness \cite{okuma2020topological, zhang2020correspondence}. Although a point with $w = 0$ is not encircled by the PBC spectrum as an oriented curve, it can be enclosed within the PBC spectrum when the spectrum is treated as a set in the $E$ plane, which divides the $w = 0$ regions into multiple connected components. This observation is crucial for linking the connectivity of the GBZ with the PBC point-gap properties, and can be summarized as a specialized version of Theorem I as follows.

{\bf Theorem II.} For a single-band model, if, for every imaginary gauge transformation of the model, the regions on the complex $E$ plane with the PBC spectrum winding number $w=0$ are connected, then the GBZ associated with the model is connected.

Here, an imaginary gauge transformation shifts the wavevector in the imaginary direction ($k\rightarrow k-i\sigma$, or equivalently, $z \rightarrow e^\sigma z$) with $\sigma$ a real parameter \cite{lee2019anatomy}, effectively replacing $t_j$ in Eq. (\ref{eq:model}) with $t_j\exp(j\sigma)$. Theorem II follows from Theorem I because the connectivity of all PBC $w=0$ regions implies a simply-connected OBC spectrum (Supplementary Materials).

The requirement in Theorem II can be satisfied by a large class of models. As a special case of Theorem II, we also prove the following.

{\bf Theorem III.} For a single-band model with a coupling range not larger than $2$ (i.e., $p\leq 2$ and $q\leq 2$), the GBZ associated with the model is connected.

Theorem III explains why disconnected GBZs do not occur in simpler, commonly-studied models. The proof shows that the PBC spectrum does not have enough complexity to completely enclose any point with $w=0$, and Theorem II applies (Supplementary Materials).

\begin{figure}
\centering
\includegraphics[width=85mm]{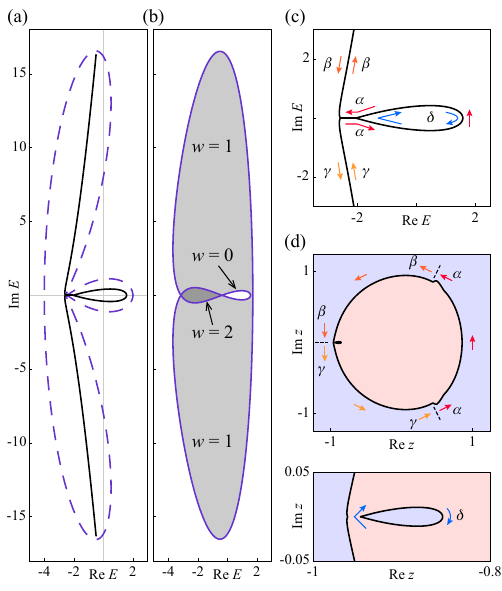}
\caption{Example of a disconnected GBZ. The model is Eq. (\ref{eq:disconn_model}) with $\epsilon = 0.5$.
(a) The PBC (dashed purple) and the OBC (solid black) spectra of the model.
(b) A gauge-transformed winding at $|z| = 0.85$, with the regions shaded according to their winding numbers. The $w=0$ region near $E = 1$ is disconnected from the outside.
(c) Enlarged view of the multiply-connected part of the OBC spectrum.
(d) GBZ of the model (top) and an enlarged view of its disconnected part (bottom). Inside and outside of the GBZ has been colored light red and light blue, respectively. Arrows with Greek letters indicate the direction for traversing the GBZ, and the corresponding path on the OBC spectrum is marked in (c).
}
\label{fig:2}
\end{figure}

{\it Examples of disconnected GBZs.} --
According to the theorems discussed above, models with disconnected GBZs must break the conditions outlined in Theorem II, where the PBC spectrum encloses a region with $w = 0$ that is disconnected from $E = \infty$. The disconnected $w=0$ regions may lead to loops in the OBC spectra and the possibility of a disconnected GBZ. 

To show the connection between the disconnected GBZ and the point-gap topology, we consider the following collection of models with $\epsilon$ a real parameter (Fig. \ref{fig:2}):
\begin{equation}
E = -z^{-3} + 4z^{-2} -4z^{-1} + 4z - 4z^2 + 3z^3 + \epsilon (z^{-1} + z - 2)
\label{eq:disconn_model}
\end{equation}
At $\epsilon = 0.5$, the model features a disconnected GBZ, with the small disconnected teardrop-shaped loop located at $z \approx -0.9$. The PBC spectrum of the model has a single connected $w=0$ region. However, the imaginary gauge transformed PBC spectrum with $|z|=0.85$ features a teardrop-shaped $w=0$ region near $E = 1$, which is disconnected from the unbounded $w=0$ region [Fig. \ref{fig:2}(b)]. The OBC spectrum also features a teardrop-shaped part and becomes multiply connected. The similarity between the three teardrop-shaped regions can be understood from the band-winding self-intersection construction of the OBC spectrum \cite{wu2022connections, wang2024non-hermitian}.

Using Eq. (\ref{eq:disconn_model}) as an example, we now discuss how some GBZ properties are modified when it becomes disconnected. When a point traverses the GBZ on the $z$ plane, its corresponding energy also traverses the OBC spectrum back and forth, covering the arcs of the OBC spectrum twice \cite{zhang2020correspondence}. This correspondence allows the Green's function of the system to be formulated as a contour integration on the GBZ \cite{hu2023greens, liu2023stable, li2022exact, zirnstein2021exponentially, chen2024formal, huang2025complex, xue2025non, xue2021simple, fu2023anatomy, mao2021boundary}. For the disconnected GBZ in Fig. \ref{fig:2}, traversing the main component of the GBZ covers most of the arcs twice, but the teardrop section on the OBC spectrum is covered only once in the counterclockwise direction [Fig. \ref{fig:2}(c)]. To cover the other direction, the disconnected component also needs to be traversed in the clockwise direction. As such, all GBZ components need to be traversed to completely cover the OBC spectrum twice.

It is also known that the GBZ loop encloses the first $p$ solutions of $z$ of $E(z) = E_0$ for any $E_0$. A more precise statement is that the first $p$ (last $q$) solutions belong to the inside (outside) of the GBZ. Note that the inside and outside notions are consistent with the winding directions of the contours, where ``inside'' (``outside'') is on the left (right) side of the contour [Fig. \ref{fig:2}(d)]. However, these assignments may be different from their actual locations in the $z$ plane, and a piece of the outside of the GBZ may be surrounded by the inside of the GBZ [Fig. \ref{fig:2}(d)].

The above two properties have been combined to show that the OBC spectrum always has a trivial point-gap topology on the $E$ plane \cite{zhang2020correspondence}. If the trivial point-gap topology refers to the property where the OBC spectrum does not divide the $E$ plane, then Fig. \ref{fig:2}(a) has provided a clear counterexample. On the other hand, the trivial point-gap topology may also refer to a zero winding number of any point on the $E$ plane when the OBC spectrum is traversed. This conclusion remains valid, provided that each arc of the OBC spectrum is traversed once in each direction. The condition is automatically satisfied by traversing all components of the GBZ in the appropriate direction that is consistent with the inside and outside designations of the regions.

\begin{figure}
\centering
\includegraphics[width=85mm]{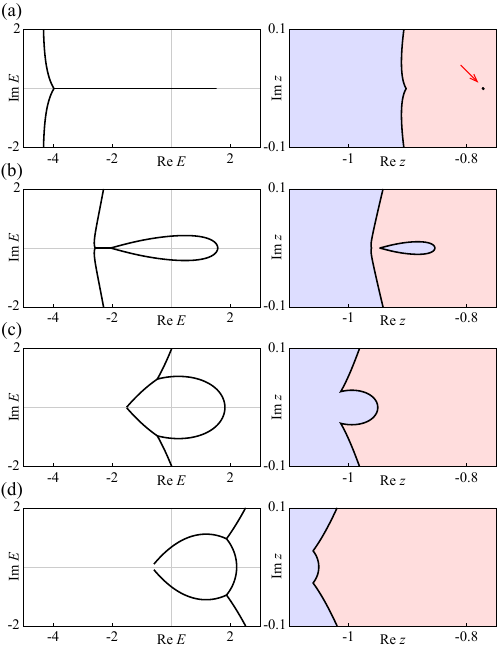}
\caption{Transitions between connected and disconnected GBZs. The models are Eq. (\ref{eq:disconn_model}) with
(a) $\epsilon = 1.4$; (b) $\epsilon = 0.5$; (c) $\epsilon = -0.5$ and (d) $\epsilon = -1.5$.
Left panels show the OBC spectra near $E = 0$ and right panels show the GBZs near $z = -1$. Inside and outside of the GBZ has been colored light red and light blue, respectively.}
\label{fig:3}
\end{figure}

Finally, we show how a disconnected GBZ can arise from a connected GBZ by changing the coupling parameters of the model. Figure \ref{fig:3} shows the OBC spectra and the GBZs for Eq. (\ref{eq:disconn_model}) with different values of $\epsilon$. The GBZ of Eq. (\ref{eq:disconn_model}) is disconnected when $0<\epsilon<1.40$. The two endpoints of this interval feature two possible transitions between a connected and a disconnected GBZ phase. Near $\epsilon\rightarrow 1.40$, the disconnected component shrinks to $z \approx -0.77$ and disappears completely when $\epsilon \approx 1.40$ [Fig. \ref{fig:3}(a)]. Near $\epsilon\rightarrow 0$, the disconnected component moves closer to the main component [Figs. \ref{fig:3}(b)], and the two components merge to form a single component at $\epsilon = 0$.

{\it Line gaps in sublattice-symmetric models.} --
Equation (\ref{eq:disconn_model}) considered above also provides an example of transitions between simply- and multiply-connected OBC spectra [Fig. \ref{fig:3}(b-d)]. We now utilize this transition to close the line gap of the OBC spectrum in a two-band model with sublattice symmetry without closing its point gap.

Models with sublattice symmetry served as prototypical examples in the study of band topologies in non-Hermitian band structures \cite{lieu2018topological, kunst2018biorthogonal, alvarez2018non-hermitian, yao2018edge, lang2018effects, kawabata2019symmetry}. The Hamiltonian of a two-band model with sublattice symmetry is given by a $2 \times 2$ matrix with only off-diagonal elements:
\begin{equation}
H = \begin{bmatrix} 0 & H_+(z) \\ H_-(z) & 0 \end{bmatrix}
\label{eq:SLS_generic}
\end{equation}
The sublattice symmetry reads $\sigma_z H \sigma_z = -H$, where $\sigma_z = \text{diag}(1,-1)$ is the third Pauli matrix. The band structure of Eq. (\ref{eq:SLS_generic}) can be found as $E^2=H_+(z)H_-(z)$, and the band energies are centrally symmetric with respect to $E = 0$ in the $E$ plane.

Generic Hermitian two-band models with sublattice symmetry possess a band gap at $E = 0$ that separates the two bands \cite{fu2023anatomy, zhou2025topological, yao2018edge}. The eigenvector topology of such a model when the gap is open is characterized by an integer topological invariant $N$. This invariant can be calculated as the integration of the Berry curvature in a single band, and $|N|$ describes the number of pairs of topological edge states at $E = 0$. This generalizes to non-Hermitian models when the point gap at $E = 0$ remains open, and the Berry curvature integration is carried out on the GBZ \cite{yao2018edge, yokomizo2019non-bloch, verma2024topological, yang2020non-hermitian}. However, as will be demonstrated below, two non-Hermitian OBC bands may merge with each other without closing the point gap, even when the band energies possess central symmetry. This is characterized by the OBC line gap of the model and is a different concept from the OBC point gap.

\begin{figure*}
\centering
\includegraphics[width=170mm]{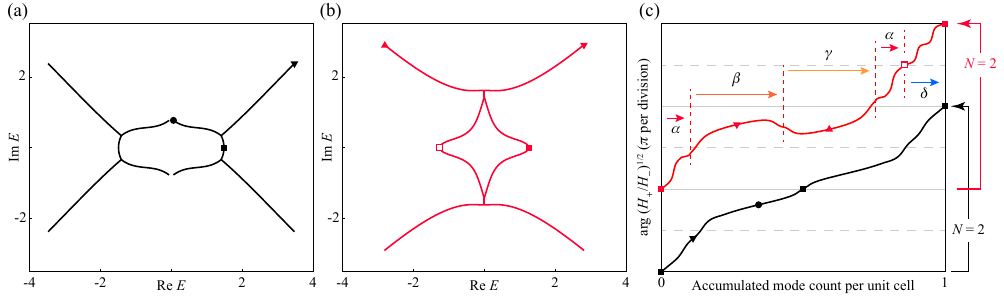}
\caption{Line-gap transitions in sublattice-symmetric models.
(a) The OBC spectra of Eq. (\ref{eq:disconn_twoband}) with $\epsilon = -1.5$ [the square root of Fig. \ref{fig:3}(d)] possesses a line gap ($\text{Re}\ E =0$).
(b) The OBC spectra of Eq. (\ref{eq:disconn_twoband}) with $\epsilon = 0.5$ [the square root of Fig. \ref{fig:3}(b)] does not possess a line gap.
(c) Berry phase $\arg \sqrt{H_+/H_-}$ along the integration contour for the two models. Markers indicate the corresponding states in (a) and (b). Greek letters mark the contour segments for (b) and are the same as in Fig. \ref{fig:2}(d).
}
\label{fig:4}
\end{figure*}

We now replace $E$ with $E^2$ in Eq. (\ref{eq:disconn_model}) and construct a two-band model with the following band structure:
\begin{equation}
E^2 = -z^{-3} + 4z^{-2} -4z^{-1} + 4z - 4z^2 + 3z^3 + \epsilon (z^{-1} + z - 2)
\label{eq:disconn_twoband}
\end{equation}
The OBC spectrum of Eq. (\ref{eq:disconn_twoband}) (referred to as ``the two-band spectrum'' below) consists of the $\pm \sqrt{E_\text{single}}$ values \cite{longhi2019probing}, where $E_\text{single}$ belongs to the OBC spectrum of Eq. (\ref{eq:disconn_model}) (``the single-band spectrum''). When the single-band spectrum is simply connected [Fig. \ref{fig:3}(d)], it is possible to draw a branch cut of the square-root function from $E = 0$ to $E = \infty$ without crossing the single-band spectrum (e.g., along the negative real axis). As a result, the two-band spectrum has two disconnected bands [Fig. \ref{fig:4}(a)]. When the single-band spectrum has a loop that contains $E = 0$ [Fig. \ref{fig:3}(b)], the square root of this loop cannot be placed within a single branch and must connect to the other branch of the same loop. The two-band spectrum therefore also has a loop that encloses $E = 0$ [Fig. \ref{fig:4}(b)]. With an increasing $\epsilon$ that induces the transition of the single-band spectrum from being simply to multiply connected, the two bands of the two-band spectrum merge together to form a single component, closing the line gap ($\text{Re}\ E =0$) without touching $E = 0$.

The absence of a line gap in sublattice-symmetric models also affects the calculation of its topological invariant. In conventional sublattice-symmetric models, the topological invariant for the number of edge states reads \cite{yao2018edge}
\begin{equation}
N = \frac{1}{2\pi}\int_\text{GBZ} d \arg \sqrt{\frac{H_+(z)}{H_-(z)}}
\end{equation}
where the contour is the GBZ for one of the two bands in the two-band spectrum. However, for models without line gaps, it is not immediately clear how the bands should be separated. Here, we propose to replace the GBZ contour in the definition of $W$ by the GBZ for the single-band spectrum (including all connected components), without regard to its corresponding band in the two-band spectrum. This substitution is always possible due to the GBZ equivalence between the two-band spectrum and the single-band spectrum \cite{longhi2019probing}.

We take Eq. (\ref{eq:disconn_twoband}) as an example and realize it as a sublattice-symmetric model. The right-hand side of Eq. (\ref{eq:disconn_twoband}) can be factored as a product of $z-z_j$ and $z^{-1}$ factors for each specific $\epsilon$, with $z_j$ the $z$ solutions of Eq. (\ref{eq:disconn_model}) at $E = 0$. These factors can then be assigned to $H_+(z)$ and $H_-(z)$ such that $E^2 = H_+(z)H_-(z)$. Different assignments share the same two-band spectra, but the eigenvector topologies can be different. The rest of the discussion uses the following specific realization:
\begin{align}
H_+ &= \sqrt{3}(z-z_1)(z-z_2)(z-z_3)/z \\
H_- &= \sqrt{3}(z-z_4)(z-z_5)(z-z_6)/z^2
\end{align}
where $z_j$ with $1 \leq j \leq 6$ as functions of $\epsilon$ are analytic continuations of the solutions of $E^2(z, \epsilon) = 0$ at $\epsilon = 0$. The accumulated Berry phase along the contour is plotted in Fig. \ref{fig:4}(c). For the case $\epsilon < 0$ with a line gap in the two-band spectrum, a conventional calculation leads to $N = 2$, which is the total Berry phase divided by $2\pi$. For the case $\epsilon > 0$ without line gaps, starting from a state with energy $E_0$ in the two-band spectrum, moving around the main component of the GBZ once will move the state to $-E_0$ instead of returning it to its original position. The main component thus contributes a half-integer ($3/2)$ to $N$. By including the contribution of the disconnected loop, which is also a half-integer ($1/2$), the total contribution becomes an integer ($N = 2$). In both cases, the invariants agree with the number of edge states found by numerically solving a system with 200 unit cells, annd also agrees with alternative methods based on ordering the zeros of $H_+$ and $H_-$ by their amplitudes \cite{lee2019anatomy, zhong2024pole}. Note that since the point gap remains open, the invariant does not change when the line gap is closed.

{\it Conclusion.} --
The connectedness of the GBZ is not universal for non-Hermitian band structures. Examples of disconnected GBZs in one-dimensional models have been demonstrated, and their connections to the point-gap properties have been discussed. Although the discussions focused mainly on specific examples, it should be clear that GBZ disconnectedness is a generic feature rather than a critical point in the parameter space. With sufficiently large coupling ranges ($p$ and $q$), it is possible to realize a GBZ with an arbitrary number of connected components. The GBZ for multiband models usually has $r$ components with $r$ the number of bands, and each piece of GBZ is associated with a single OBC band \cite{yang2020non-hermitian}. As such, the disconnected GBZ concept can be generalized to a GBZ possessing more connected components than the number of bands. The models with such disconnected GBZs can be created by coupling single-band chains together, with each chain having a disconnected single-band GBZ. These examples of disconnected GBZ in multi-band models contrast earlier assumptions that the GBZ for each band is a contour that encircle the $z=0$ origin in a counterclockwise manner \cite{yokomizo2020non-blochb, verma2024topological, yang2020non-hermitian, guo2021nonhermitian, yokomizo2020topological, fu2023anatomy}.

Our results point to the necessity of reexamining existing arguments that depend on the GBZ topology. The Berry curvature calculation provides such an example, where the interpretation of a single band becomes ambiguous, and the integration contours for the topological invariant \cite{yokomizo2019non-bloch, yao2018edge, verma2024topological, zhongwang2025topological, yang2020non-hermitian, imura2019generalized, hou2022deterministic, verma2024non, liu2023topological} or the Green's function \cite{hu2023greens, li2022exact, zirnstein2021exponentially, chen2024formal, huang2025complex, xue2025non, xue2021simple, fu2023anatomy, mao2021boundary} need to be chosen carefully. Another example is the braiding characteristics of the OBC spectrum \cite{fu2024braiding, wang2025observing, li2022topological, shi2024studies}. With a disconnected GBZ, it is not immediately clear how the GBZ should be traversed in a single pass to obtain the energy strands. Overall, we believe that our results on the connectivity of the GBZ point to new directions concerning the bands and gaps in non-Hermitian systems and their topological properties in general.

This work is funded by a Simons Investigator in Physics grant from the Simons Foundation (Grant No. 827065).

\clearpage
\onecolumngrid
\appendix

\section*{Supplementary Materials}

\subsection{Definitions}

Before stating and proving the theorems in the main text, we formalize some definitions that we will use below.

A {\it single-band model}, or a {\it model} for short, is a Laurent polynomial in $z$ given by $E = \sum_{-p\leq j \leq q} t_jz^j$, where $t_j$ are complex numbers, $p \geq 1$ and $q \geq 1$ are positive integers representing the coupling ranges, $t_{-p}\neq 0$ and $t_{q}\neq 0$. Equivalent forms of the model include:
\begin{equation}
- E + \sum_{-p\leq j \leq q} t_jz^j = 0
\end{equation}
and
\begin{equation}
- E z^p + \sum_{0\leq j \leq p+q} t_{j-p}z^j = 0
\end{equation}
The {\it PBC spectrum} is the range of the model as a mapping from $z$ to $E$ when restricted to $|z|=1$.

A {\it solution} of the model at a specific $E$ is a $z$ value that satisfies the model when the value for $E$ has been substituted. By the fundamental theorem of algebra, there are always $p+q$ solutions for a specific $E$, counted with multiplicity. We order these solutions by their absolute value in non-decreasing order, such that $z_j(E)$ refers to the solution with the $j$th smallest absolute value. The $E$ dependence may be dropped if the value of $E$ is clear from the context. For completeness, at $E = \infty$ we define $z_j = 0$ when $1 \leq j \leq p$ and $z_j = \infty$ when $p+1 \leq j \leq p+q$. By the continuity of polynomial roots, $|z_j|$ is a continuous function of $E$ for each $j$, although $z_j$ is not always continuous because the ordering of the roots may change when $|z_j|=|z_{j+1}|$.

\begin{figure*}
\centering
\includegraphics[width=170mm]{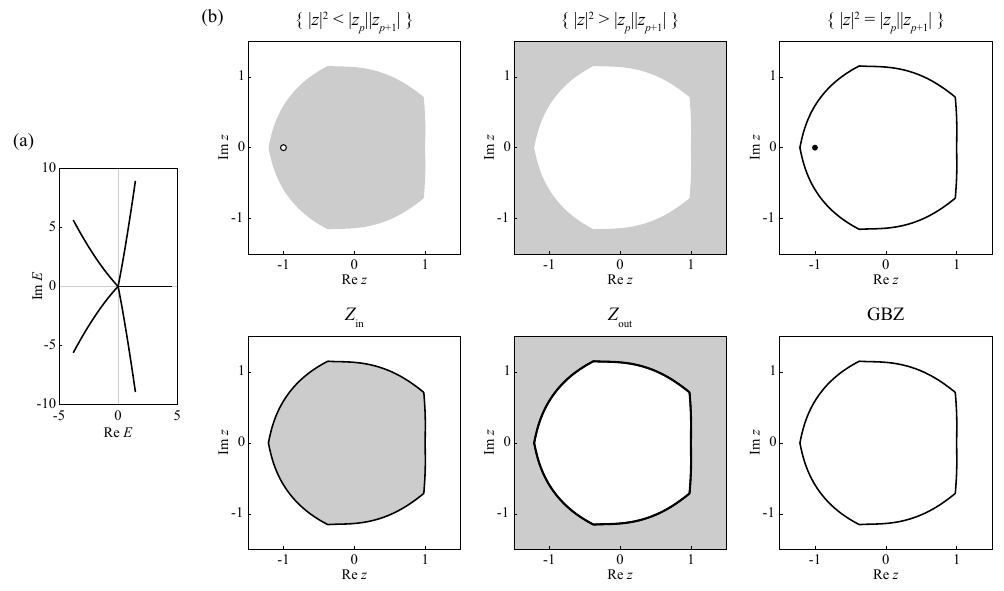}
\caption{
(a) The OBC spectrum of the model $E = -2z^{-3} + (3+\sqrt{5})z^{-2} + (-3+\sqrt{5})z^2 +2z^3$.
(b) Different sets of $z$ defined for the model $E = -2z^{-3} + (3+\sqrt{5})z^{-2} + (-3+\sqrt{5})z^2 +2z^3$. Black curves indicate that the boundary is included in the sets, and hollow circles indicate the points are missing from the sets.}
\label{fig:S1}
\end{figure*}

The {\it inside of the GBZ}, denoted as $Z_\text{in}$, is defined as the closure of the set $\{z: |z|^2 < |z_{p}(E)| |z_{p+1}(E)|\}$ on the Riemann sphere of $z$, where $E = E(z)$. Similarly, the {\it outside of the GBZ}, denoted as $Z_\text{out}$, is defined as the closure of the set $\{z: |z|^2 > |z_{p}(E)| |z_{p+1}(E)|\}$ on the Riemann sphere of $z$. Examples of thee sets can be found in Fig. \ref{fig:S1}. The closure operations include the region boundaries in the sets, so $Z_\text{in}$ becomes a subset of $\{|z|^2 \leq |z_{p}| |z_{p+1}|\}$ and $Z_\text{out}$ becomes a subset of $\{|z|^2 \geq |z_{p}| |z_{p+1}|\}$. For later convenience, we include $z=0$ in $\{|z|^2 < |z_{p}||z_{p+1}|\}$ and $z=\infty$ in $\{|z|^2 > |z_{p}||z_{p+1}|\}$ using the continuity of polynomial roots. When $z$ is not $0$ or $\infty$, $E(z)$ is well-defined, and $|z|^2$ can be smaller, larger, or equal to $|z_{p}| |z_{p+1}|$. Since $z$ must be among the solutions of the model when $E = E(z)$, $|z|^2 < |z_{p}||z_{p+1}|$ implies that $|z| \leq |z_{p}(E)|$, and $z$ is among the first $p$ solutions of the model for $E$ that corresponds to $z$. Roughly speaking, $Z_\text{in}$ contains the states that are relatively more localized to the left. The case $|z|^2 > |z_{p}||z_{p+1}|$ implies that $|z| \geq |z_{p+1}(E)|$, and $z$ is among the last $q$ solutions of the model. $Z_\text{out}$ thus contains the states that are relatively more localized to the right. Finally, the case $|z|^2 = |z_{p}||z_{p+1}|$ implies $|z| = |z_{p}(E)| = |z_{p+1}(E)|$, which is the condition for the existence of the OBC states.

The {\it GBZ} is defined as the set $Z_\text{in}\cap Z_\text{out}$. Clearly, the GBZ defined this way is a subset of $\{z: |z_p| = |z_{p+1}|\}$, but it might differ from $\{z: |z_p| = |z_{p+1}|\}$ by isolated points. The closure operations in the definition of $Z_\text{in}$ and $Z_\text{out}$ ensure that these isolated $|z_p|=|z_{p+1}|$ points are not included in the GBZ but on the same sides of the GBZ as their neighborhoods (Fig. \ref{fig:S1}).

The {\it OBC spectrum} is the set $\{E: |z_p| = |z_{p+1}|\}$. It is the image of the set $\{z: |z_p| = |z_{p+1}|\}$ under the mapping from $z$ to $E$ given by the model. However, since the OBC spectrum does not contain isolated points, the image of any isolated point in $\{E: |z_p| = |z_{p+1}|\}$ will be covered by the non-isolated sections, and the OBC spectrum can also be characterized as the image of the GBZ under the mapping from $z$ to $E$.

\subsection{Proof of Theorem I}

{\bf Theorem I.} For a single-band model, if the OBC spectrum of the model is simply connected, then the GBZ associated with the model is connected.

We note that the exclusion of isolated points from the GBZ is necessary because such points can appear for models with simply-connected OBC spectrum and are disconnected from the GBZ (Fig. \ref{fig:S1}).

{\bf Proof.} We first observe that the complement of the OBC spectrum on the $E$ Riemann sphere must be connected. In fact, if the complement has at least two connected components, there must be at least one component not containing $E = \infty$ and this component is bounded. By construction, the OBC spectrum encloses this component, which produces a nontrivial loop in the OBC spectrum and contradicts its simple connectedness. This shows that any $E$ point not on the OBC spectrum is path-connected to $E = \infty$.

Now consider a point $z\neq 0$ that satisfies $|z|^2 < |z_{p}(E)| |z_{p+1}(E)|$ where $E = E(z)$. By definition, this point belongs to the interior of $Z_\text{in}$. The corresponding $E(z)$ does not belong to the OBC spectrum. As a result, there is a continuous path on the $E$ Riemann sphere that connects $E(z)$ to $E = \infty$ without intersecting the OBC spectrum. By the continuity of polynomial roots, there is a corresponding continuous path on the $z$ Riemann sphere that connects $z$ to $z=0$ (i.e., the path can be lifted from $E$ to $z$). Since the path on the $z$ Riemann sphere does not pass through $\{|z_p| = |z_{p+1}|\}$, the condition $|z|^2 < |z_{p}| |z_{p+1}|$ is maintained. Therefore, the set $\{|z|^2 < |z_{p}| |z_{p+1}|\}$ is connected.

We next show that the interior of $Z_\text{in}$, which is an open set, is connected. In addition to $\{|z|^2 < |z_{p}| |z_{p+1}|\}$, the interior of $Z_\text{in}$ may also contain isolated $|z_p|=|z_{p+1}|$ points that are completely surrounded by $\{|z|^2 < |z_{p}| |z_{p+1}|\}$. A path connecting such isolated points to $z=0$ can be constructed by first connecting it to a point in its neighborhood that satisfies $|z|^2 < |z_{p}| |z_{p+1}|$ and then connecting to $z=0$. This composite path remains in the interior of $Z_\text{in}$, validating its connectivity.

The connectivity of $Z_\text{in}$, which is a closed set, can be similarly established. The boundary of $Z_\text{in}$ consists of a subset of $\{|z_p| = |z_{p+1}|\}$, where the neighborhood of each point contains both $|z|^2 < |z_{p}| |z_{p+1}|$ and $|z|^2 > |z_{p}| |z_{p+1}|$ points. For each such point, a path within $Z_\text{in}$ can be constructed that connects it to $z=0$.

The same argument shows that the interior of $Z_\text{out}$ is connected and $Z_\text{out}$ is connected, replacing $|z|^2 < |z_{p}| |z_{p+1}|$ by $|z|^2 > |z_{p}| |z_{p+1}|$ and $z=0$ by $z=\infty$.

We then show that $Z_\text{in}\cup Z_\text{out}$ is the entire Riemann sphere of $z$, that is, each point in $\{|z_p| = |z_{p+1}|\}$ belongs to $Z_\text{in}$ or $Z_\text{out}$. We note that the set $\{|z_{p}(E)| = |z_{p+1}(E)|\}$ is a subset of the zeros of a nontrivial polynomial of $\text{Re}\ z$ and $\text{Im}\ z$. As such, no point in $\{|z_p| = |z_{p+1}|\}$ can have a neighborhood that lies entirely in $\{|z_p| = |z_{p+1}|\}$, as a generic line through this point intersects $\{|z_p| = |z_{p+1}|\}$ at infinitely many points, contradicting the fundamental theorem of algebra. Consequently, any neighborhood of a point in $\{|z_p| = |z_{p+1}|\}$ must contain dense subsets of $\{|z|^2 < |z_{p}| |z_{p+1}|\}$, $\{|z|^2 > |z_{p}| |z_{p+1}|\}$, or both. These points are included in the closure of $\{|z|^2 < |z_{p}| |z_{p+1}|\}$, $\{|z|^2 > |z_{p}| |z_{p+1}|\}$, or both, respectively.

As $Z_\text{in}$ and $Z_\text{out}$ cover the Riemann sphere of $z$ and have no common interior points, the GBZ becomes the common boundary between $Z_\text{in}$ and $Z_\text{out}$. The complement of $Z_\text{out}$ is therefore the interior of $Z_\text{in}$, and vice versa. A theorem concerning the topology of the Riemann sphere states that an open set is simply connected if both the set and its complement in the Riemann sphere are connected \cite{conway1978functions}. Applying this to the interior of $Z_\text{in}$ shows that the interior of $Z_\text{in}$ is simply connected. By the Riemann mapping theorem, the interior of $Z_\text{in}$ is topologically equivalent to a unit disk, and the GBZ as its boundary is connected.

\subsection{Proof of Theorem II}

{\bf Theorem II.} For a single-band model, if, for every imaginary gauge transformation of the model, the regions on the complex $E$ plane with the PBC spectrum winding number $w=0$ are connected, then the GBZ associated with the model is connected.

We note that a connected $w=0$ region on the complex $E$ plane is equivalent to a connected $w=0$ region on the $E$ Riemann sphere, since a tight-binding PBC spectrum is bounded and the point $E = \infty$ always has a winding number of $0$. In addition, because winding numbers are not defined on the PBC spectrum, two disjoint $w=0$ regions with touching corners or edges are considered disconnected regions.

{\bf Proof.} We show that the prerequisite condition in Theorem II implies that the OBC spectrum is simply connected, and Theorem I applies. This is done using proof-by-contradiction.

Assume that the OBC spectrum of the single-band model is multiply connected and choose a nontrivial loop of the OBC spectrum. Choose a $E = E_0$ point in the enclosed region of the loop that does not belong to the OBC spectrum. The solutions to the model at this $E_0$ feature $|z_{p}| < |z_{p+1}|$. As such, an imaginary transformation in the form of $z' = z/\sqrt{|z_{p}||z_{p+1}|}$ features exactly $p$ solutions within the unit circle of $z'$, and the transformed PBC spectrum has a winding number of $0$ at $E_0$. In addition, the point $E = \infty$ always has a winding number of $0$ for a bounded PBC spectrum.

We now observe that the complement of the OBC spectrum has at least two connected components, with $E_0$ and $E = \infty$ belonging to different connected components that are separated by the nontrivial loop of the OBC spectrum. Since the OBC spectrum is a subset of the union of the $w \neq 0$ regions and the PBC spectrum \cite{bottcher1999introduction, okuma2020topological}, the $w=0$ regions are subsets of the complements of the OBC spectrum. Two disjoint components remain disjoint after taking subsets, and the $E_0$ and $E = \infty$ points belong to different connected components of $w=0$. This indicates that the $w=0$ regions are not connected, contradicting the prerequisite that the $w=0$ regions are connected. Therefore, a connected $w=0$ for every imaginary gauge transformation implies a simply connected OBC spectrum, and Theorem II follows from Theorem I.

\subsection{Proof of Theorem III}

{\bf Theorem III.} For a single-band model with a coupling range not larger than $2$ (i.e., $p\leq 2$ and $q\leq 2$), the GBZ associated with the model is connected.

{\bf Proof.} We show that the prerequisite condition in Theorem III implies that the PBC spectrum does not divide the $w=0$ region. Since imaginary gauge transformations do not change the coupling range of the model, the connectedness of the $w=0$ region holds for all imaginary gauge transformations, and Theorem II applies.

We first consider the case $p=q=2$ and explicitly represent the model as follows:
\begin{equation}
E = t_{-2}z^{-2} + t_{-1}z^{-1} + t_0 + t_1z+ t_2z^2
\end{equation}
with $t_{-2}\neq 0$ and $t_2\neq 0$. Consider a point $E = E_0$ that has winding number $w=0$ with respect to the PBC spectrum. The winding number indicates that two solutions at $E_0$ are inside the unit circle and two solutions are outside ($|z_1| \leq |z_2| < 1 < |z_3| \leq |z_4|$). By the fundamental theorem of algebra, we can factor the model as follows:
\begin{equation}
E - E_0= t_2\frac{(z-z_1)(z-z_4)}{z}\frac{(z-z_2)(z-z_3)}{z}
\end{equation}

We now study the factor $(z-z_1)(z-z_4)/z$. When $|z| = 1$, the image of $(z-z_1)(z-z_4)/z$ is an ellipse, as seen by writing $z = \exp(ik)$ and expanding the real and imaginary parts in terms of trigonometric functions of $k$. In addition, since $|z_1| < 1 < |z_4|$, the winding number of the ellipse with respect to the origin is zero, and the origin is outside the ellipse. The convexity of the ellipse implies that the angle subtended by the ellipse with respect to the origin is strictly less than $\pi$. The same argument shows that the angle subtended by $(z-z_2)(z-z_3)/z$ with respect to the origin is also strictly less than $\pi$.

The PBC spectrum relative to $E_0$ is the product of a constant and two factors dependent on  $z$. The argument of a product is the sum of the arguments of each factor, and the range of the argument is not larger than the range of arguments of each factor combined. As such, the subtended angle of the PBC spectrum with respect to $E_0$ is strictly smaller than $2\pi$. Therefore, it is possible to find a ray originating from $E_0$ that extends to $E = \infty$ without intersecting the PBC spectrum. As the choice of $E_0$ is arbitrary, this shows that the entire $w=0$ region is connected, and Theorem III follows from Theorem II.

The case $p=1$ and $q=2$ can be obtained as a limit of $t_{-2} \rightarrow 0$ and $z_1 \rightarrow 0$. One of the factors becomes a circle, which remains convex, and the upper bound of the subtended angle is not affected. As such, the $w=0$ regions remain connected. The same argument can be applied to the case $p=2$ and $q=1$. Therefore, the GBZ is also connected for these cases.

It is possible to prove the case $p=q=1$ by also taking limits. However, we note that this case will reduce the model to a Hatano-Nelson model, where the GBZ is always a circle and connected.

\end{document}